# Two distinct superconducting states controlled by orientation of local wrinkles in LiFeAs

Lu Cao[1,2,*], Wenyao Liu[1,2,*], Geng Li[1,2,3,4,*,†], Guangyang Dai[1,2], Qi Zheng[1,2], Kun Jiang[1,2], Shiyu Zhu[1,2], Li Huang[1,2,3], Lingyuan Kong[1], Fazhi Yang[1,2], Xiancheng Wang[1,3,4], Wu Zhou[2,3], Xiao Lin[2,1], Jiangping Hu[1,2,3], Changqing Jin[1,3,4], Hong Ding[1,3,4,‡], Hong-Jun Gao[1,2,3,4,§]

[1]Institute of Physics, Chinese Academy of Sciences, Beijing 100190, China

[2]School of Physical Sciences, University of Chinese Academy of Sciences, Beijing 100049, China

[3]CAS Center for Excellence in Topological Quantum Computation, University of Chinese Academy of Sciences, Beijing 100190, China

[4]Songshan Lake Materials Laboratory, Dongguan, Guangdong 523808, China

**We observe two types of superconducting states controlled by orientations of local wrinkles on the surface of LiFeAs. Using scanning tunneling microscopy/spectroscopy, we find type-I wrinkles enlarge the superconducting gaps and enhance the transition temperature, whereas type-II wrinkles significantly suppress the superconducting gaps. The vortices on wrinkles show a $C_2$ symmetry, indicating the strain effects on the wrinkles. A discontinuous switch of superconductivity occurs at the border between two different wrinkles. Our results demonstrate that the local strain effect could affect superconducting order parameter of LiFeAs with a possible Lifshitz transition, by alternating crystal structure in different directions.**

The origin of superconductivity in iron-based superconductors (FeSCs) remains elusive despite intensive research efforts over a dozen years [1,2]. The large orbital degrees of freedom as well as the presence of intertwining orders hinder a microscopic understanding of the pairing mechanism in FeSCs [3,4]. As a perturbation method, external pressure can lift the ground state degeneracy and offer detailed information about how the unconventional superconductivity evolve with other electronic orders. For example, an in-plane resistivity anisotropy [5] and an spin excitation [6] have been observed in electron-doped $BaFe_2As_2$ under uniaxial pressure; in FeSe with external pressure, the magnetic order could emerge and coexist with the high-temperature superconductivity [7], while enhanced spin fluctuation is evidenced as well [8]. Among FeSCs, LiFeAs is unique as its phase diagram is not intervened by any magnetic or nematic order, which provides an appropriate platform to detect the relationship between pressure and superconductivity in strong-correlation system [9,10].

Using scanning tunneling microscopy/spectroscopy (STM/S), we report the observation of orientation-dependent superconductivity at two types of wrinkle on LiFeAs surface. The tunneling spectra show an increase of superconducting (SC) gaps on type-I wrinkles and a reduction on type-II wrinkles. Temperature-dependent measurements of the SC gap show that the gap closing temperature on type-I wrinkles is enhanced by 20 ~ 30 %, but it remains almost unchanged on type-II wrinkles. While wrinkles are a commonly consequence of relieving transverse strain induced by the change of atom coordination [11,12], the spatial feature of superconducting vortices observed on wrinkles confirms the exist of local strains. Our results demonstrate that strains with different orientations can enhance or suppress superconductivity of LiFeAs, offering new insights into the SC mechanism of FeSCs.

The atomic model of LiFeAs is shown in Fig. 1(a). Unlike many other FeSCs [13], the cleavage of LiFeAs crystal occurs at a non-polar plane between the two Li layers [dashed line in Fig. 1(a)], presenting a good platform for investigating and tuning the unconventional superconductivity at nanoscale [14-17]. The stoichiometric LiFeAs shows superconductivity below transition temperature $T_c$ (~17 K) [18]. Two types of wrinkles are observed on LiFeAs surface [Figs. 1(b)-1(d)], appearing as straight 1D ridges. Type-I wrinkles extend along the [110] direction (with respect to Li surface, also the Fe-Fe direction) or its neighboring directions, spanning a width about 15 nm with a maximum height of ~1.0 Å [lower panel, Fig. 1(c)]. Type-II wrinkles extend along the [100] direction (also the Fe-As direction) or its neighboring directions, with a width of ~10 nm and a maximum height of ~0.7 Å [lower panel, Fig. 1(d)]. Both types of wrinkles are uniform in width and extend from several tens of to hundreds of nanometers. Atomic resolution image of type-I wrinkle in Fig. 1(e) shows continuous and perfect Li lattice, excluding the possibility of formation of twin boundary [19,20], domain wall [21] or line defects [22]. Within the resolution of STM, no obvious lattice constant change can be detected. We propose that these wrinkles are likely induced by releasing of local strain during the creation of LiFeAs surface upon cleavage. Indeed, by atomic force microscopy (AFM), we explicitly demonstrate that the wrinkles on LiFeAs surface have real spatial corrugations, instead of purely reflecting as enhancement of local density of states (LDOS) (Supplemental Material Fig. S1 [23]).

Here we find the remarkable difference between type-I and II wrinkles lies in their LDOS. Figure 1(f) displays the differential conductance spectra (d$I$/d$V$) taken at the wrinkle-free region (black curve), type-I (red curve) and type-II wrinkles (blue curve) on the surface, as marked by the crosses in Figs. 1(c) and 1(d). In the wrinkle-free region, d$I$/d$V$ spectrum shows multiband features of LiFeAs, with a large gap of ~5.8 meV (may come from the inter-hole pocket at the Γ point) and a small gap of ~2.9 meV (may come from the outer-hole pocket at the Γ point), consistent with previous reports [14-16,24]. Interestingly, the tunneling spectrum on type-I wrinkle yields a coherence peak at 7.3 meV and a shoulder at 3.6 meV, and type-II wrinkle has a single V-shaped gap of 2.5 meV. We note that electron doping [17] and application of external pressure [9,10] normally lead to reduction of SC gap or $T_c$, and that increase of SC gap is not commonly seen in LiFeAs.

Inspired by the novel gap features observed on the wrinkles, we carry out spatial d$I$/d$V$ spectra line-cut across and along the wrinkles, as displayed in Figs. 2(a)-2(d). On type-I wrinkle, the coherence peaks of $\Delta_1$ start to shift to higher energies when getting close to the wrinkle edge and show constant values of ±7.3 meV across the wrinkle [Fig. 2(a)]. At the same time, the shoulders of $\Delta_2$ follow a similar tendency. The enlarged gaps $\Delta_1$ and $\Delta_2$ remain homogeneous along the type-I wrinkle [Fig. 2(b)]. Furthermore, the gap map of a type-I wrinkle (Supplemental Material Figs. S2a,b [23]) reveals the maximum gap size ($\Delta_1$) distribution. It is evident that the type-I wrinkle has larger SC gap sizes compared with the wrinkle-free region. Also, the edges of the wrinkle (Supplemental Material Figs. S2a,b [23]) show the largest SC gap, possibly suggesting the highest strain at the edges. We compare the SC gaps between wrinkle-free region and type-I region under different axis scales (Supplemental Material Fig. S2c [23]). The results show that the enhancement of the two SC gaps is of the same ratio (20~30 %). On type-II wrinkle, however, the coherence peaks of $\Delta_1$ are strongly suppressed when getting close to the wrinkle edge and totally disappear on the wrinkle region, meanwhile, the shoulder of $\Delta_2$ evolves into a pair of coherence peaks on the wrinkle [Fig. 2(c)]. Note that the gap features are robust and homogenous along type-II wrinkle [Fig. 2(d)].

We next perform the temperature-dependent d$I$/d$V$ measurements. On type-I wrinkle, the SC gap $\Delta_1$ can be well differentiated at 17 K ($T_c$ of bulk LiFeAs) and gradually closes at ~20.5 K [Fig. 3(a)], while the gap closes at 17 K at wrinkle-free region [Fig. 3(b)]. The SC gap of type-II wrinkle also closes at ~17 K [Fig. 3(c)]. We note that there is bump near $E_F$ at high temperature (black rows in Fig. 3(a)-3(b)), which should suggest the band top of $d_{xz}$ [30]. In Fig. 3(d), we plot the extracted gap values as a function of temperature. The wrinkle-free region (black squares) and type-I region (red triangles) follow the same tendency, which is more obvious after rescaling the data of type-I region (pink triangles). This coincidence suggests that type-I region has similar coupling strength with that of the wrinkle-free region. However, on the type-II regions, superconductivity behaves differently. Compared with its gap size, such $T_c$ on type-II wrinkle is quite high and maybe come from the bulk superconducting proximity effect while the intrinsic superconductivity on type-II wrinkle is already suppressed.

To explain the exotic superconducting behaviors we observed, the first step is understanding the circumstances on these wrinkles. Here, we image the vortex structure by zero-bias conductance (ZBC) map (large scale map in Supplemental Material Fig. S4 [23]). The vortex on wrinkle-free region present four-pointed-star like structure due to the C$_4$ symmetry of the Fermi surface [Fig. 4(a)], in contrast, the vortex at wrinkles change from C$_4$ symmetric to C$_2$ symmetric shape with the long axis extending along the wrinkle orientation [Fig. 4(b)]. We point out that our vortex structure results not only clarify the existence and orientation of local uniaxial strain on the wrinkles, but also suggests that the configuration of Fermi surface changes consequently [25].

Next, we plot the statistics of the orientation of the wrinkles and the corresponding SC gaps, which reveals that the [110] direction favors the type-I wrinkles while the [100] direction favors the type-II wrinkles. Interestingly, wrinkles along other orientations are also found in our experiment [Fig. 4(c)], whereas the gap size versus orientation yields an abrupt change at ~20° with respect to the [110] direction. The statistical results suggest that the SC gaps do not gradually evolve with wrinkle orientations, but resemble a discontinuous transition [Fig. 4(c)]. We also note that the wrinkles can make turns on the surface, leading to a transition from type-I to type-II (Supplemental Material Fig. S5 [23]). The transition is smooth without observable boundaries.

Further analysis indicates that the switch of the superconductivity on two-type wrinkles is coincidence with the band structure changing under different orientations of local strain. First, the bump features in Figs. 3(a)-(b) indicate the band of $d_{xz}$ shifts above the Fermi level on the type-I wrinkles. Next, our results indicate that the wrinkles impact the LDOS away from the Fermi level as well [Fig. 4(d)]. Note that there is a shoulder around 30 meV which has a slight move comparing to the one in wrinkle-free region (Supplemental Material Fig. S3 [23]), suggesting the band top of the $d_{xy}$ shifts up at the type-I wrinkle and down at the type-II wrinkle [30]. In addition, the whole gap feature changes from U-shape to V-shape [17] with non-zero density of states at the zero energy on the type-II wrinkle [Fig. 1(f), Fig. 3(c) and Fig. 4(d)], and the intensity of LDOS for type-II has a large loss compare to others as well, both indicate a possibility that the inter-hole band sinks down to the Fermi level on the type-II region.

Based on the observations discussed above, we propose a possible scenario that the local strain on wrinkle cause a Lifshitz transition [26-28]. In bulk LiFeAs, the $d_{xz}/d_{yz}$ band splits possibly due to the spin-orbit coupling [29], with $d_{yz}$ crossing $E_F$ and $d_{xz}$ sinking below [Fig. 4(f)]. When local strain is present along the [110] direction (Fe-Fe direction), the $d_{xz}$ band top shifts above $E_F$, giving rise to an increase of DOS near Γ point [Fig. 4(e)]. The enhanced DOS increase the SC gap and $T_c$. The same increasing ratio of $\Delta_1$ and $\Delta_2$ on type-I wrinkle also supports this scenario. On the other hand, once the strain is along [100] direction (Fe-As direction), both the $d_{yz}$ and $d_{xz}$ band sink below $E_F$, leaving only the $d_{xy}$ band crossing $E_F$ [Fig. 4(g)]. In this case, the Lifshitz transition of Fermi surface leads to the disappearance of the $\Delta_1$ gap, and that only the $\Delta_2$ gap can be observed, yet the V-shape of the in-gap states is quite puzzling. This V-shape SC gap resemble the tunneling spectra in electron-doped LiFe$_{1-x}$Co$_x$As [17]. We also note that since the superconductivity in LiFeAs might relates to spin fluctuations [8], it is possible that As anion height from Fe atoms [30] which will be affected by the local strain may modify the spin fluctuations and consequently affects the superconductivity. Nevertheless, we cannot rule out other scenarios of superconducting changing and further theoretical understanding of mechanism is required.

In summary, we have identified two types of wrinkles on LiFeAs surface. The orientation dependent wrinkles accompanied with local uniaxial strains have very different influence on the superconductivity at the nanoscale, with one get an enhancement while the other is suppressed. A Lifshitz transition scenario is proposed to explain these two distinct states of superconductivity in the vicinity of wrinkles on LiFeAs. Our observations suggest that the change of band structure has strong influence for unconventional superconductivity in LiFeAs.

We thank Hu Miao and Fuchun Zhang for helpful discussion. This work is supported by the Ministry of Science and Technology of China (2019YFA0308500, 2018YFA0305700, 2016YFA0401000), the National Natural Science Foundation of China (11888101, 52072401, 61888102, 51991340, 11820101003,11921004, 11674371), and the Chinese Academy of Sciences (XDB28000000, XDB07000000).

*These authors contributed equally to this work.
†Corresponding author.
gengli.iop@iphy.ac.cn
‡Corresponding author.
dingh@iphy.ac.cn
§Corresponding author.
hjgao@iphy.ac.cn

# References

[1] G. R. Stewart, Rev. Mod. Phys. **83**, 1589 (2011).

[2] J. Paglione and R. L. Greene, Nat. Phys. **6**, 645 (2010).

[3] R. M. Fernandes, A. V. Chubukov, and J. Schmalian, Nat. Phys. **10**, 97 (2014).

[4] Q. Si, R. Yu, and E. Abrahams, Nat. Rev. Mater. **1**, 16017 (2016).

[5] J. H. Chu, J. G. Analytis, K. De Greve, P. L. McMahon, Z. Islam, Y. Yamamoto, and I. R. Fisher, Science **329**, 824 (2010).

[6] X. Y. Lu, J. T. Park, R. Zhang, H. Q. Luo, A. H. Nevidomskyy, Q. M. Si, and P. C. Dai, Science **345**, 657 (2014).

[7] J. P. Sun *et al.*, Nat. Commun. **7**, 12146 (2016).

[8] J. P. Sun *et al.*, Phys. Rev. Lett. **118**, 147004 (2017).

[9] X. C. Wang, S. J. Zhang, Q. Q. Liu, Z. Deng, Y. X. Lv, J. L. Zhu, S. M. Feng, and C. Q. Jin, High Pressure Res. **31**, 7 (2011).

[10] C. M. Yim, C. Trainer, R. Aluru, S. Chi, W. N. Hardy, R. Liang, D. Bonn, and P. Wahl, Nat. Commun. **9**, 2602 (2018).

[11] D. Edelberg, H. Kumar, V. Shenoy, H. Ochoa, and A. N. Pasupathy, Nat. Phys. **16**, 1097 (2020).

[12] J. Mao *et al.*, Nature **584**, 215 (2020).

[13] L. Cao *et al.*, accept by Nano Res. (2021).

[14] T. Hanaguri, K. Kitagawa, K. Matsubayashi, Y. Mazaki, Y. Uwatoko, and H. Takagi, Phys. Rev. B **85**, 214505 (2012).

[15] S. Chi, S. Grothe, R. X. Liang, P. Dosanjh, W. N. Hardy, S. A. Burke, D. A. Bonn, and Y. Pennec, Phys. Rev. Lett. **109**, 087002 (2012).

[16] M. P. Allan *et al.*, Science **336**, 563 (2012).

[17] J.-X. Yin *et al.*, Phys. Rev. Lett. **123**, 217004 (2019).

[18] X. C. Wang, Q. Q. Liu, Y. X. Lv, W. B. Gao, L. X. Yang, R. C. Yu, F. Y. Li, and C. Q. Jin, Solid State Commun. **148**, 538 (2008).

[19] C. L. Song, Y. L. Wang, Y. P. Jiang, L. Wang, K. He, X. Chen, J. E. Hoffman, X. C. Ma, and Q. K. Xue, Phys. Rev. Lett. **109**, 137004 (2012).

[20] T. Watashige *et al.*, Phys. Rev. X. **5**, 031022 (2015).

[21] Z. Y. Wang, J. O. Rodriguez, L. Jiao, S. Howard, G. M., G. D. Gu, T. L. Hughes, D. K. Morr, and V. Madhavan, Science **367**, 104 (2020).

[22] C. Chen, K. Jiang, Y. Zhang, C. Liu, Y. Liu, Z. Wang, and J. Wang, Nat. Phys. **16**, 536 (2020).

[23] See Supplemental Material.

[24] K. Umezawa *et al.*, Phys. Rev. Lett. **108**, 037002 (2012).

[25] V. Sunko *et al.*, npj Quantum Mater. **4**, 46 (2019).

[26] I. M. Lifshitz, Sov. Phys. JEPT **11**, 1130 (1960).

[27] C. Liu *et al.*, Nat. Phys. **6**, 419 (2010).

[28] X. Shi *et al.*, Nat. Commun. **8**, 14988 (2017).

[29] H. Miao *et al.*, Phys. Rev. B **89**, 220503(R) (2014).

[30] Y. Mizuguchi, Y. Hara, K. Deguchi, S. Tsuda, T. Yamaguchi, K. Takeda, H. Kotegawa, H. Tou, and Y. Takano, Supercond. Sci. Technol. **23**, 054013 (2010).

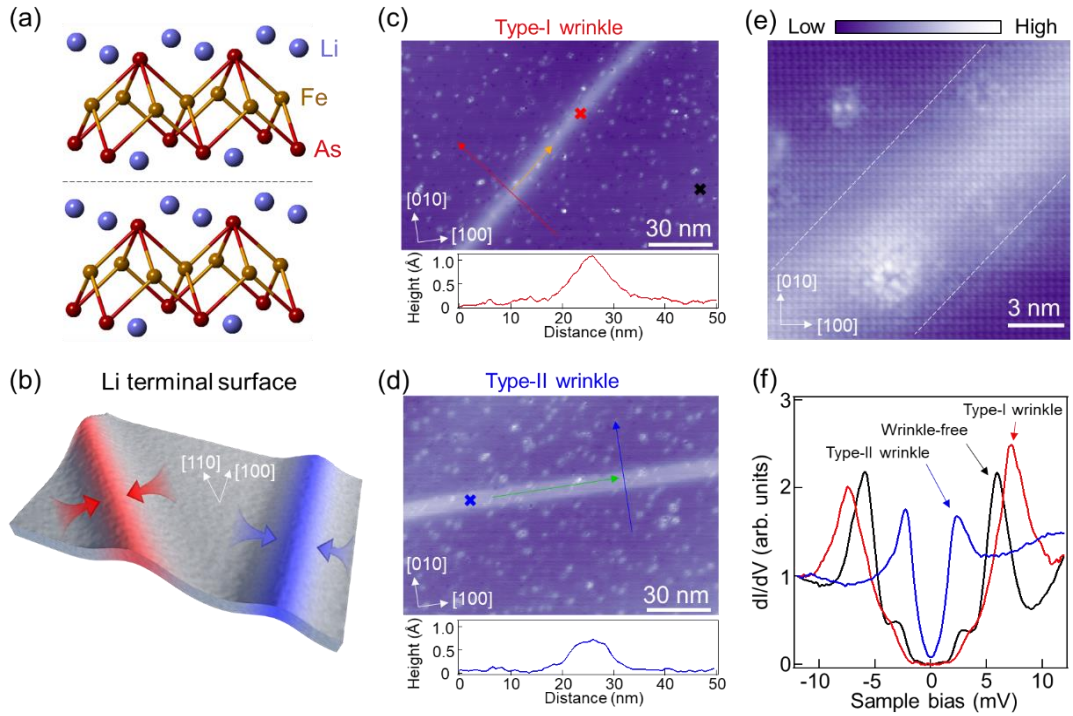

FIG. 1. (a) The crystal structure of LiFeAs. The black dashed line indicates the location where the cleavage happens. (b) The sketches of the two types of wrinkles and the local strain on LiFeAs surface. (c) Upper panel: a large scale STM topography of LiFeAs surface showing a type-I wrinkle. The [100] and [010] mark the lattice directions on the Li terminal surface, respectively. Setpoint: $V_s$ = -20 mV, $I_t$ = -20 pA. Lower panel: a line profile taken along the red line in the upper panel. (d) Same of (c) but showing a type-II wrinkle. (e) The atomic resolution image of type-I wrinkle. The white dashed line indicates the wrinkle edge. Setpoint: $V_s$ = -3 mV, $I_t$ = -1 nA. (f) The d$I$/d$V$ spectra taken at the three crosses in (c) and (d). Setpoint: $V_s$ = -10 mV, $I_t$ = -200 pA.

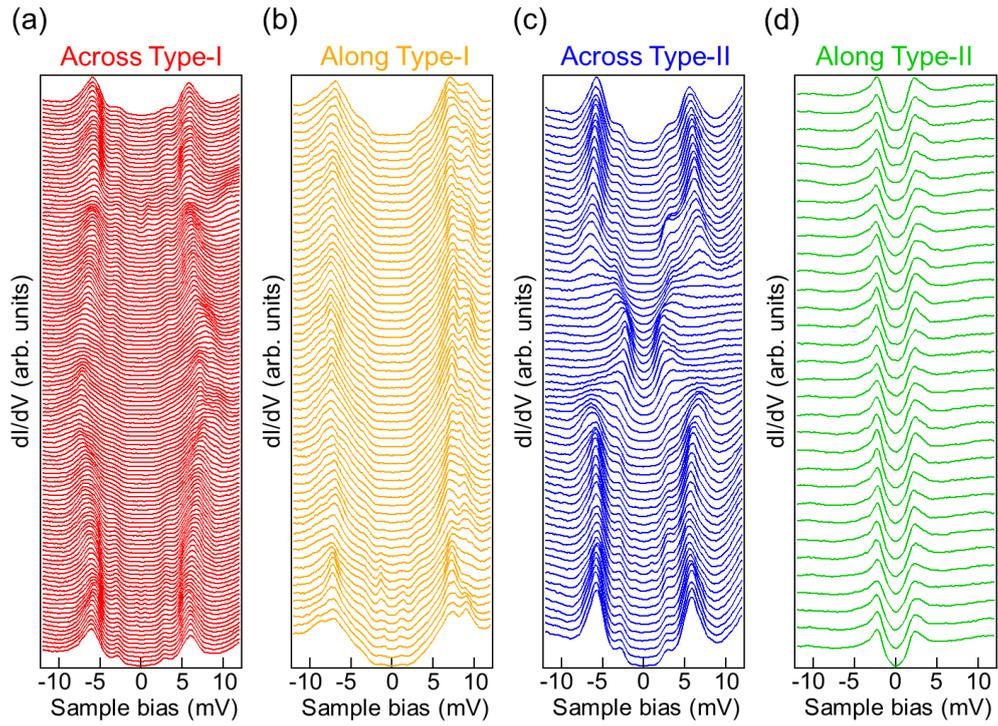

FIG. 2. (a)-(b) The waterfall plots of d$I$/d$V$ spectra across (red) and along (orange) the type-I wrinkle as marked by the arrows in Fig. 1(c), showing the homogeneous and enlarged superconducting gap along type-I wrinkle. (c)-(d) The waterfall plots of d$I$/d$V$ spectra across (blue) and along (green) the type-II wrinkle as marked by the arrows in Fig. 1(d), showing the homogeneous and reduced superconducting gap along type-II wrinkle. Setpoint: $V_s$ = -10 mV, $I_t$ = -200 pA.

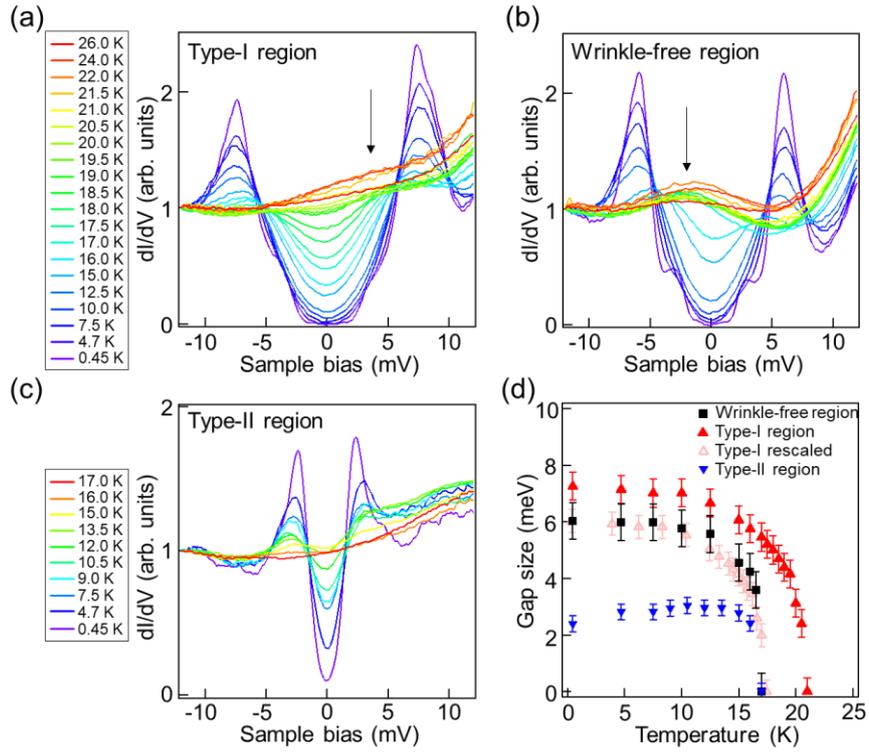

FIG. 3. (a)-(c) Temperature dependent d$I$/d$V$ spectra taken at type-I wrinkle region, wrinkle-free region and type-II wrinkle region, respectively. The black arrows highlight the bump features of the LDOS. (a) and (b) share the same legend. Setpoint: $V_s$ = -10 mV, $I_t$ = -200 pA. (d) A plot of the gap sizes as a function of temperature extracted from (a)-(c).

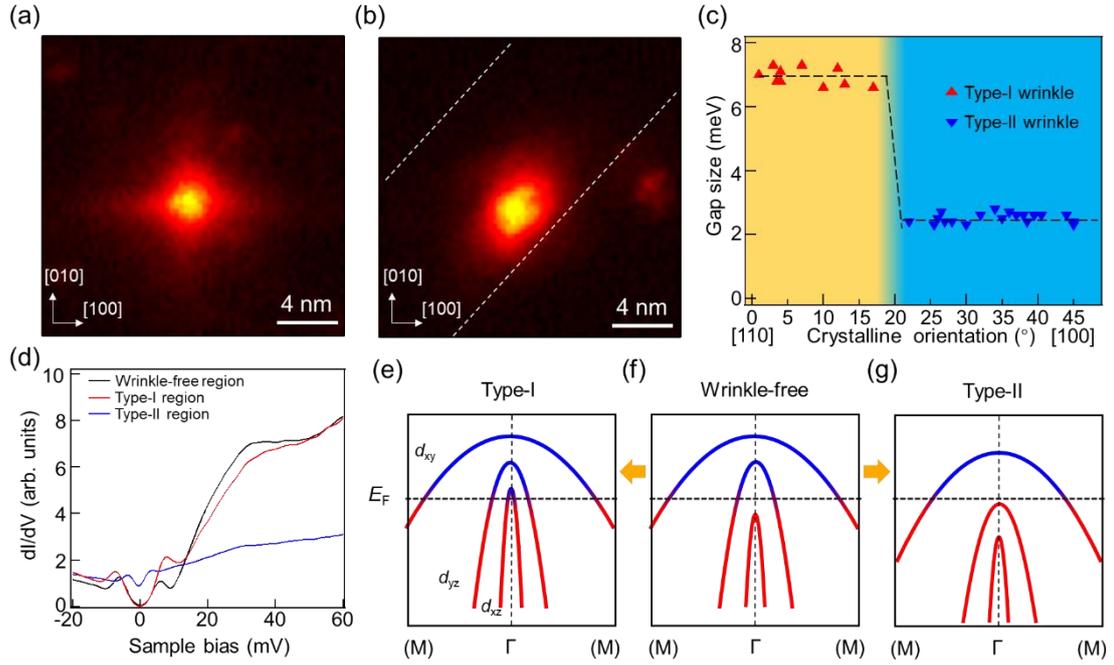

FIG. 4. (a)-(b) Zero-bias conductance map of vortices at wrinkle-free region and type-I wrinkle, respectively. The white dashed line in (b) indicate the wrinkle edge. (c) Statistics of orientations of the two types of wrinkles. The dashed line is a guide to eye. (d) The wide range of d$I$/d$V$ spectra. (e)-(g) Sketches of the band structures of LiFeAs near the $\Gamma$ point and the bands shift of two types of wrinkles.

# Supplemental Materials

## Methods

**Single-crystal growth and scanning tunneling microscopy/spectroscopy (STM/S)**

High-quality LiFeAs single crystals were synthesized by using the self-flux method [31]. The LiFeAs crystals were mounted at the STM sample holder by epoxy in the groove box and transferred to the ultra-high vacuum chamber where they were cleaved *in-situ*, and then immediately transferred to the STM scanner at 4.7 K. The STM/S measurements were operated at 0.45 K by $^3$He single-shot technique (Unisoku). Polycrystal tungsten tips were etched chemically and calibrated on Au(111) surface before use. All STM images were acquired in the constant-current mode. The differential conductance (d$I$/d$V$) spectra and maps were obtained by a standard lock-in amplifier at a frequency of 973.0 Hz, with a modulation voltage of 0.2 mV. Magnetic fields were applied perpendicular to the samples.

**Non-contact atomic force microscopy (nc-AFM)**

The nc-AFM measurements were conducted at liquid He temperature with a base ultra-high vacuum lower than 2 × 10$^{-10}$ mbar, where the samples were cleaved *in-situ*. A commercial qPlus tuning fork sensor in frequency modulation mode with Pt/Ir tip was used to obtain the data. The resonance frequency was about 27.9 kHz and the stiffness was about 1800 N/m. The STM topography images were acquired in constant-current mode. The constant-high and constant-force AFM modes were used to measure the real topography features of the two types of wrinkles.

## The real topography of wrinkles

In STM, it is well known that the tunneling current is proportional to the integral of the density of states (DOS) from 0 to the setpoint bias. So the topography obtained from the constant-current mode is the contour of the constant DOS, which strictly speaking is not the real topography. To clarify the wrinkle structure observed by STM has a real hump on topography rather than a result from the high DOS accumulation or charge accumulation, we carry out the atomic force microscopy (AFM) measurements. Unlike the STM, AFM acquires the atomic force between the tip and the sample, thus can be used to detect the real topography. We first use the constant-current STM mode to find the type-I and II wrinkles as well as the wrinkle-free region (Fig. S1a,b). Then, the constant-high AFM maps are used to detect the atomic force on the wrinkles and wrinkle-free region (Fig. S1c,d). Note that the stronger atomic force (dark color in Fig. S1c,d) means the shorter tip-sample distance, therefore the wrinkles have a real hump topography due to the constant-high mode. We also use the constant-force AFM mode to check the wrinkles (Fig. S1e,f). In the constant-force mode, the tip-sample distance is tuned to keep the atomic force keep constant. So the absolute height of topography can be detected. The results of the constant-force line profile shown in Fig. S1e,f demonstrate again that the wrinkles have a real hump topography.

## The wide range d$I$/d$V$ spectra on wrinkles

The hump feature at ~30 meV corresponds to the band top of the $d_{xy}$ band, which shows a slight shift at different regions. The second differential curves (Fig. S3a) of the data further indicates the local band ($d_{xy}$) shift on the wrinkles. We find that the $d_{xy}$ band shifts upward at type-I region and downward at type-II region.

## The vortex lattice on wrinkles

Although the type-I and II wrinkles have different orientations, the vortices on wrinkles present a lattice structure with equal distance (Fig. S4b,d). More importantly, compared with the vortex shape on wrinkle-free regions, they all break the $C_4$ symmetry with an oval-like shape with the long axis along the wrinkles, which strongly indicates not only that the local strain exists at the wrinkles, but also that the local electronic state is dramatically changed.

## The transition between type-I and type-II wrinkles

We find a kink that divide the type-I and type-II wrinkles (Fig. S5a), and the d$I$/d$V$ spectra from the type-II wrinkles (Fig. S5b) to type-I wrinkles (Fig. S5c) exhibit the smooth and continuous SC gaps transition from a single gap feature with 2.6 meV to the two enlarged gap features with 7.0 meV and 4.0 meV.

[31]  L. Y. Xing, H. Miao, X. C. Wang, J. Ma, Q. Q. Liu, Z. Deng, H. Ding, and C. Q. Jin, J. Phys.: Condens. Matter **26**, 435703 (2014).

**Supplemental data**

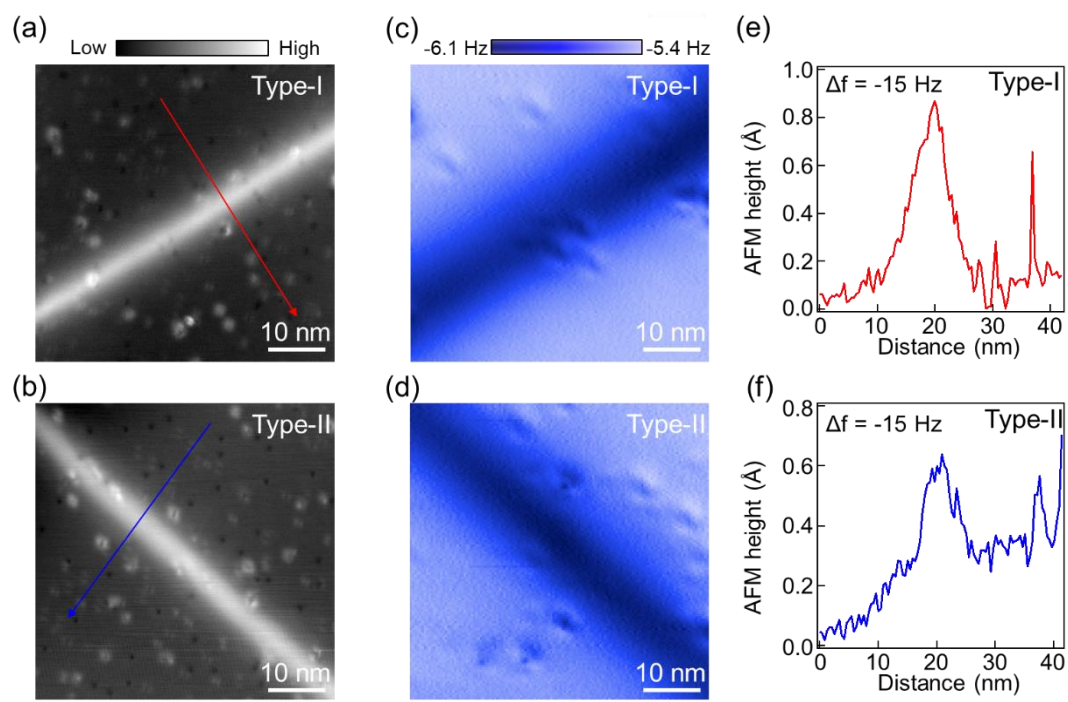

FIG. S1. (a)-(b) STM topographies of type-I and type-II wrinkles, respectively. (c)-(d) Constant-height AFM maps on the two types of wrinkles. The bright and dark color scale correspond to the weak and strong tip-sample interactions, respectively, indicating that the two types of wrinkles have real spatial corrugations on the LiFeAs surface. (e)-(f) Constant-force AFM line-profile measured along the arrows marked in (a) and (b).

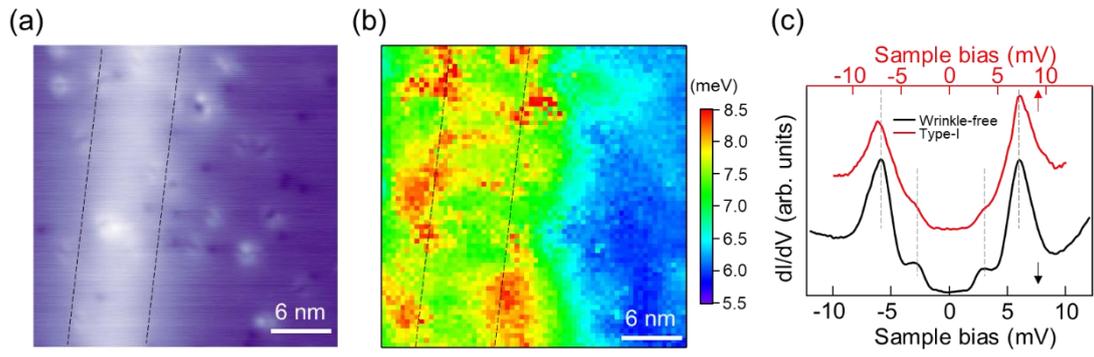

FIG. S2. (a) A zoom-in STM topography of a type-I wrinkle. Black dashed lines indicate the edges of the wrinkle. (b) A gap map of $\Delta_1$ taken at a type-I wrinkle. (c) Comparison of the SC gaps between the wrinkle-free region and type-I region in different axis scales. The grey dashed lines mark the two coherence peaks. For type-I region, the two gaps ($\Delta_1$ and $\Delta_2$) increase at an equal ratio.

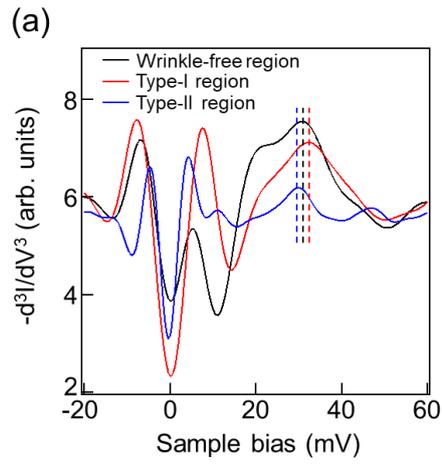

FIG. S3. (a) The 2$^{nd}$ differential curves of the spectra in Fig. 4(a). The dashed lines indicate the energy positions of the $d_{xy}$ band top.

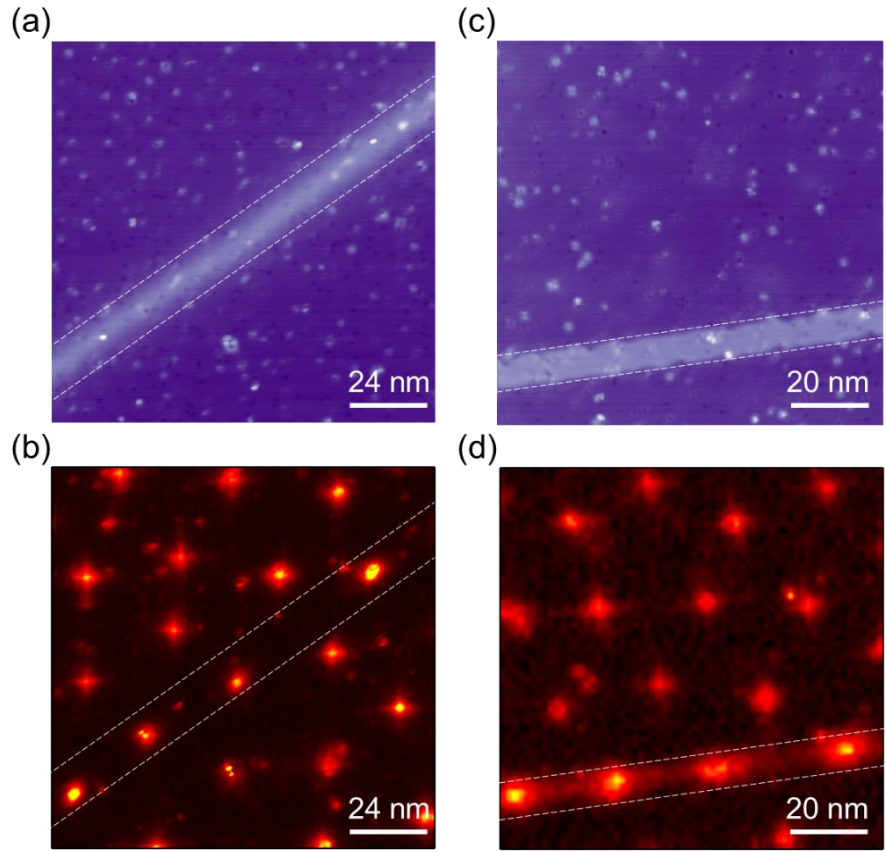

FIG. S4. (a) A large scale STM topography image with wrinkle-free region and type-I wrinkle region included. White dashed lines mark the edge of wrinkle. (b) The zero-bias conductance (ZBC) map of the same area in (a) under a magnetic field of 3 T, showing the vortices distributed in the whole area. (c)-(d) Same as (a) and (b), but with type-II wrinkle region included and under the magnetic field of 4 T.

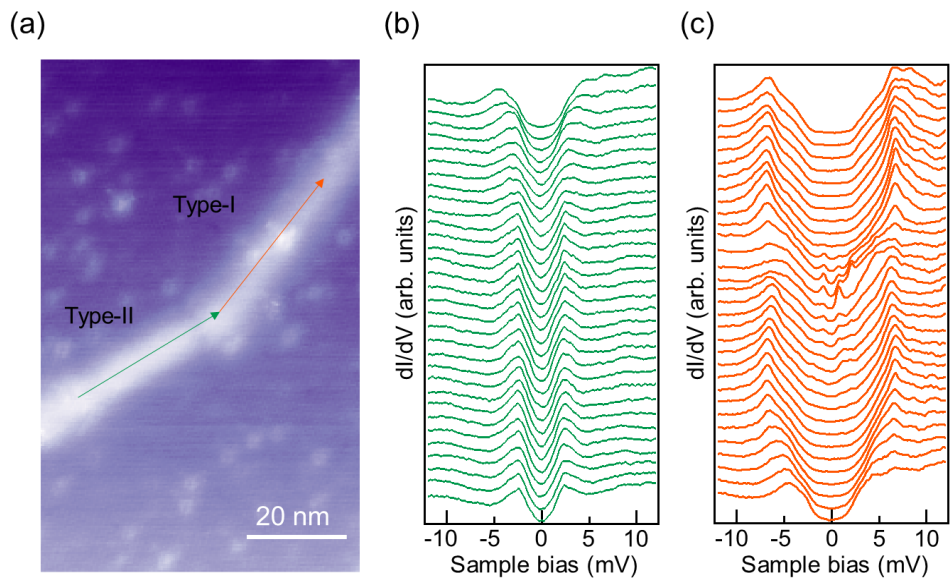

FIG. S5. (a) A STM topography showing the kink which divides the type-I and type-II wrinkles. (b)-(c), The d$I$/d$V$ spectra taken along the arrows marked in (a), showing the SC gap transition from type-II to type-I wrinkles smoothly.